\documentclass[12pt,a4paper]{article}
\usepackage[utf8]{inputenc}
\usepackage{amsmath,amsfonts,amssymb}
\usepackage{graphicx}
\usepackage{cite}
\usepackage{algorithm}
\usepackage{algorithmic}
\usepackage{url}
\usepackage{hyperref}
\usepackage{booktabs}
\usepackage{array}
\usepackage{subfigure}

\title{Direction of Arrival Estimation: A Tutorial Survey of Classical and Modern Methods}

\author{
	Amgad A. Salama \\
	\textit{The Research and Development Center, ADC, Cairo, Egypt} \\
	amgad.salama@acm.org
}

\date{}

\begin{document}

\maketitle

\begin{abstract}
Direction of arrival (DOA) estimation is a fundamental problem in array signal processing with applications spanning radar, sonar, wireless communications, and acoustic signal processing. This tutorial survey provides a comprehensive introduction to classical and modern DOA estimation methods, specifically designed for students and researchers new to the field. We focus on narrowband signal processing using uniform linear arrays, presenting step-by-step mathematical derivations with geometric intuition. The survey covers classical beamforming methods, subspace-based techniques (MUSIC, ESPRIT), maximum likelihood approaches, and sparse signal processing methods. Each method is accompanied by Python implementations available in an open-source repository, enabling reproducible research and hands-on learning. Through systematic performance comparisons across various scenarios, we provide practical guidelines for method selection and parameter tuning. This work aims to bridge the gap between theoretical foundations and practical implementation, making DOA estimation accessible to beginners while serving as a comprehensive reference for the field. See \url{https://github.com/AmgadSalama/DOA} for detail implementation of the methods.
\end{abstract}

\section{Introduction}

Direction of arrival (DOA) estimation is the process of determining the spatial directions from which signals impinge on an array of sensors. This fundamental problem in array signal processing has been extensively studied for over five decades, driven by diverse applications in radar target tracking, sonar navigation, wireless communications beamforming, acoustic source localization, and seismic monitoring.

The problem can be stated simply: given measurements from an array of spatially separated sensors, estimate the angles from which multiple source signals arrive. While conceptually straightforward, DOA estimation presents numerous challenges including limited array aperture, finite sample size, noise corruption, model uncertainties, and computational constraints.

\subsection{Historical Perspective}

The evolution of DOA estimation methods reflects advances in signal processing theory and computational capabilities. Early approaches in the 1960s relied on classical beamforming techniques, essentially steering the array in different directions and measuring the output power. The 1970s saw the development of adaptive beamforming methods like Capon's minimum variance distortionless response (MVDR) beamformer.

A major breakthrough came in the 1980s with the introduction of subspace-based methods, particularly the multiple signal classification (MUSIC) algorithm by Schmidt and the estimation of signal parameters via rotational invariance techniques (ESPRIT) by Roy and Kailath. These methods achieved super-resolution capability, overcoming the fundamental limitations of classical beamforming.

The 1990s brought maximum likelihood (ML) approaches, providing optimal performance at the cost of increased computational complexity. More recently, the 2000s and 2010s have seen the emergence of sparse signal processing techniques, leveraging compressed sensing theory to achieve high-resolution DOA estimation.

\subsection{Survey Scope and Contributions}

This tutorial survey differs from existing reviews in several key aspects:

\begin{itemize}
\item \textbf{Educational focus}: We prioritize clarity and accessibility over comprehensive coverage, providing detailed mathematical derivations with geometric intuition.
\item \textbf{Narrowband emphasis}: By focusing exclusively on narrowband signals, we avoid the additional complexity of wideband processing while covering the fundamental concepts.
\item \textbf{Uniform linear arrays}: We restrict attention to uniform linear arrays (ULA), the most common and well-understood array geometry.
\item \textbf{Implementation-oriented}: Each method is accompanied by clean Python implementations in an open-source repository.
\item \textbf{Systematic comparison}: We provide comprehensive performance analysis across standardized test scenarios.
\end{itemize}

The main contributions of this work include:
\begin{enumerate}
\item A unified mathematical framework for DOA estimation methods
\item Step-by-step derivations with clear geometric interpretation
\item Open-source Python implementations of all discussed methods
\item Systematic performance comparison and practical guidelines
\item A structured learning path for students entering the field
\end{enumerate}

\subsection{Organization}

The remainder of this paper is organized as follows. Section 2 establishes the mathematical foundation and signal model. Sections 3-6 present the main classes of DOA estimation methods: classical beamforming, subspace-based techniques, maximum likelihood approaches, and sparse signal processing methods. Section 7 covers specialized techniques for challenging scenarios. Section 8 provides comprehensive performance analysis and comparison. Section 9 discusses practical implementation considerations. Section 10 describes the accompanying software repository. Section 11 concludes with a summary and future directions.

\section{Problem Formulation and Signal Model}

\subsection{Array Geometry}

We consider a uniform linear array (ULA) consisting of $M$ omnidirectional sensors with inter-element spacing $d$. The sensors are positioned along the $x$-axis at locations $\mathbf{p}_m = [(m-1)d, 0, 0]^T$ for $m = 1, 2, \ldots, M$.

The ULA offers several advantages for educational purposes:
\begin{itemize}
\item Simple steering vector expressions
\item Well-understood beampattern characteristics  
\item Enables polynomial rooting methods (Root-MUSIC)
\item Most widely used in practice
\end{itemize}

\subsection{Signal Model}

Consider $K$ narrowband far-field sources located at angles $\{\theta_1, \theta_2, \ldots, \theta_K\}$ relative to the array normal (broadside direction). The received signal at the $m$-th sensor can be written as:

\begin{equation}
x_m(t) = \sum_{k=1}^{K} s_k(t) e^{j\omega \tau_{m,k}} + n_m(t)
\end{equation}

where $s_k(t)$ is the complex envelope of the $k$-th source signal, $\omega$ is the angular frequency, $\tau_{m,k}$ is the propagation delay from the $k$-th source to the $m$-th sensor, and $n_m(t)$ is additive noise.

For a far-field source at angle $\theta_k$, the propagation delay to the $m$-th sensor is:

\begin{equation}
\tau_{m,k} = \frac{(m-1)d \sin(\theta_k)}{c}
\end{equation}

where $c$ is the propagation velocity.

Collecting all sensor outputs into a vector $\mathbf{x}(t) = [x_1(t), x_2(t), \ldots, x_M(t)]^T$, we obtain the fundamental array signal model:

\begin{equation}
\mathbf{x}(t) = \mathbf{A}\mathbf{s}(t) + \mathbf{n}(t)
\label{eq:signal_model}
\end{equation}

where:
\begin{itemize}
\item $\mathbf{s}(t) = [s_1(t), s_2(t), \ldots, s_K(t)]^T$ is the source signal vector
\item $\mathbf{n}(t) = [n_1(t), n_2(t), \ldots, n_M(t)]^T$ is the noise vector
\item $\mathbf{A} = [\mathbf{a}(\theta_1), \mathbf{a}(\theta_2), \ldots, \mathbf{a}(\theta_K)]$ is the $M \times K$ array manifold matrix
\end{itemize}

The steering vector for a source at angle $\theta$ is given by:

\begin{equation}
\mathbf{a}(\theta) = [1, e^{j\omega\tau_1}, e^{j\omega\tau_2}, \ldots, e^{j\omega\tau_{M-1}}]^T
\end{equation}

For a ULA with half-wavelength spacing ($d = \lambda/2$), this simplifies to:

\begin{equation}
\mathbf{a}(\theta) = [1, e^{j\pi\sin\theta}, e^{j2\pi\sin\theta}, \ldots, e^{j(M-1)\pi\sin\theta}]^T
\label{eq:steering_vector}
\end{equation}

\subsection{Statistical Assumptions}

Throughout this survey, we make the following standard assumptions:

\begin{enumerate}
\item \textbf{Far-field assumption}: Sources are located in the far-field, allowing plane wave approximation
\item \textbf{Narrowband assumption}: Source signals have bandwidth much smaller than the reciprocal of the maximum propagation delay across the array
\item \textbf{Additive noise}: The noise $\mathbf{n}(t)$ is additive, zero-mean, and uncorrelated with source signals
\item \textbf{Array calibration}: The array geometry and sensor characteristics are perfectly known
\end{enumerate}

Additional assumptions specific to certain methods will be stated as needed.

\subsection{Covariance Matrix}

Many DOA estimation methods operate on the array covariance matrix rather than the instantaneous measurements. The theoretical covariance matrix is defined as:

\begin{equation}
\mathbf{R} = E[\mathbf{x}(t)\mathbf{x}^H(t)] = \mathbf{A}\mathbf{R}_s\mathbf{A}^H + \sigma^2\mathbf{I}
\label{eq:covariance}
\end{equation}

where $\mathbf{R}_s = E[\mathbf{s}(t)\mathbf{s}^H(t)]$ is the source covariance matrix, $\sigma^2$ is the noise variance, and $\mathbf{I}$ is the identity matrix.

In practice, the covariance matrix is estimated using $N$ temporal snapshots:

\begin{equation}
\hat{\mathbf{R}} = \frac{1}{N}\sum_{n=1}^{N}\mathbf{x}[n]\mathbf{x}^H[n]
\end{equation}

The quality of this estimate depends on the number of snapshots $N$ and affects the performance of all covariance-based methods.

\section{Classical Beamforming Methods}

Classical beamforming methods form the foundation of array signal processing. These techniques are intuitive, robust, and computationally efficient, making them excellent starting points for understanding DOA estimation.

\subsection{Delay-and-Sum Beamforming}

The delay-and-sum (DAS) beamformer, also known as conventional beamforming, is the simplest DOA estimation method. The basic idea is to steer the array to different look directions and measure the output power.

\subsubsection{Principle}

For a look direction $\theta$, the beamformer applies phase shifts to align signals from that direction and then averages across sensors:

\begin{equation}
y(t, \theta) = \frac{1}{M}\mathbf{a}^H(\theta)\mathbf{x}(t) = \frac{1}{M}\sum_{m=1}^{M}x_m(t)e^{-j(m-1)\pi\sin\theta}
\end{equation}

The spatial spectrum is formed by computing the output power:

\begin{equation}
P_{DAS}(\theta) = E[|y(t, \theta)|^2] = \frac{1}{M^2}\mathbf{a}^H(\theta)\mathbf{R}\mathbf{a}(\theta)
\label{eq:das_spectrum}
\end{equation}

DOA estimates correspond to peaks in this spatial spectrum.

\subsubsection{Geometric Interpretation}

The DAS beamformer can be understood geometrically. When signals from direction $\theta$ are received by the array, they experience different phase shifts at each sensor. By applying conjugate phase shifts and summing, signals from the look direction add constructively while signals from other directions add incoherently.

The beampattern $B(\theta, \phi)$ shows the array response when steered to direction $\theta$ for a signal arriving from direction $\phi$:

\begin{equation}
B(\theta, \phi) = \frac{1}{M}\left|\mathbf{a}^H(\theta)\mathbf{a}(\phi)\right| = \frac{1}{M}\left|\frac{\sin(M\pi(\sin\phi - \sin\theta)/2)}{\sin(\pi(\sin\phi - \sin\theta)/2)}\right|
\end{equation}

\subsubsection{Algorithm}

\begin{algorithm}
\caption{Delay-and-Sum Beamforming}
\begin{algorithmic}[1]
\REQUIRE Array data $\mathbf{x}[n]$, $n = 1, \ldots, N$
\REQUIRE Angular grid $\theta_i$, $i = 1, \ldots, I$
\STATE Estimate covariance matrix: $\hat{\mathbf{R}} = \frac{1}{N}\sum_{n=1}^{N}\mathbf{x}[n]\mathbf{x}^H[n]$
\FOR{$i = 1$ to $I$}
    \STATE Compute steering vector: $\mathbf{a}(\theta_i)$
    \STATE Compute spatial spectrum: $P_{DAS}(\theta_i) = \frac{1}{M^2}\mathbf{a}^H(\theta_i)\hat{\mathbf{R}}\mathbf{a}(\theta_i)$
\ENDFOR
\STATE Find peaks in $P_{DAS}(\theta)$ to obtain DOA estimates
\end{algorithmic}
\end{algorithm}

\subsubsection{Advantages and Limitations}

\textbf{Advantages:}
\begin{itemize}
\item Simple implementation and intuitive operation
\item Robust to model mismatch and calibration errors
\item Stable performance across different scenarios
\item Low computational complexity
\end{itemize}

\textbf{Limitations:}
\begin{itemize}
\item Limited resolution (approximately $\lambda/L$ where $L$ is array length)
\item Poor performance in presence of strong interferers
\item Cannot resolve closely spaced sources
\item Susceptible to sidelobe interference
\end{itemize}

\subsection{Capon Beamforming}

The Capon beamformer, also known as the minimum variance distortionless response (MVDR) beamformer, improves upon conventional beamforming by adaptively nulling interfering signals while maintaining unity gain in the look direction.

\subsubsection{Principle}

The Capon beamformer solves the optimization problem:

\begin{align}
\min_{\mathbf{w}} \quad & \mathbf{w}^H\mathbf{R}\mathbf{w} \\
\text{subject to} \quad & \mathbf{w}^H\mathbf{a}(\theta) = 1
\end{align}

where $\mathbf{w}$ is the beamforming weight vector. This minimizes the output power while maintaining unity response in the look direction $\theta$.

Using Lagrange multipliers, the optimal weight vector is:

\begin{equation}
\mathbf{w}_{Capon}(\theta) = \frac{\mathbf{R}^{-1}\mathbf{a}(\theta)}{\mathbf{a}^H(\theta)\mathbf{R}^{-1}\mathbf{a}(\theta)}
\end{equation}

The corresponding spatial spectrum is:

\begin{equation}
P_{Capon}(\theta) = \frac{1}{\mathbf{a}^H(\theta)\mathbf{R}^{-1}\mathbf{a}(\theta)}
\label{eq:capon_spectrum}
\end{equation}

\subsubsection{Mathematical Derivation}

Starting from the constrained optimization problem, we form the Lagrangian:

\begin{equation}
\mathcal{L} = \mathbf{w}^H\mathbf{R}\mathbf{w} + \lambda^*(\mathbf{w}^H\mathbf{a}(\theta) - 1) + \lambda(1 - \mathbf{a}^H(\theta)\mathbf{w})
\end{equation}

Taking the derivative with respect to $\mathbf{w}^*$ and setting to zero:

\begin{equation}
\frac{\partial\mathcal{L}}{\partial\mathbf{w}^*} = \mathbf{R}\mathbf{w} + \lambda\mathbf{a}(\theta) = 0
\end{equation}

This gives $\mathbf{w} = -\lambda\mathbf{R}^{-1}\mathbf{a}(\theta)$. Substituting into the constraint:

\begin{equation}
-\lambda\mathbf{a}^H(\theta)\mathbf{R}^{-1}\mathbf{a}(\theta) = 1
\end{equation}

Solving for $\lambda$ and substituting back yields the optimal weight vector.

\subsubsection{Algorithm}

\begin{algorithm}
\caption{Capon Beamforming}
\begin{algorithmic}[1]
\REQUIRE Array data $\mathbf{x}[n]$, $n = 1, \ldots, N$
\REQUIRE Angular grid $\theta_i$, $i = 1, \ldots, I$
\STATE Estimate covariance matrix: $\hat{\mathbf{R}} = \frac{1}{N}\sum_{n=1}^{N}\mathbf{x}[n]\mathbf{x}^H[n]$
\STATE Add diagonal loading if needed: $\hat{\mathbf{R}} \leftarrow \hat{\mathbf{R}} + \epsilon\mathbf{I}$
\FOR{$i = 1$ to $I$}
    \STATE Compute steering vector: $\mathbf{a}(\theta_i)$
    \STATE Compute spatial spectrum: $P_{Capon}(\theta_i) = \frac{1}{\mathbf{a}^H(\theta_i)\hat{\mathbf{R}}^{-1}\mathbf{a}(\theta_i)}$
\ENDFOR
\STATE Find peaks in $P_{Capon}(\theta)$ to obtain DOA estimates
\end{algorithmic}
\end{algorithm}

\subsubsection{Practical Considerations}

\textbf{Diagonal Loading:} In practice, the sample covariance matrix may be ill-conditioned, especially when $N < M$. Diagonal loading helps stabilize the matrix inversion:

\begin{equation}
\hat{\mathbf{R}}_{loaded} = \hat{\mathbf{R}} + \epsilon\mathbf{I}
\end{equation}

where $\epsilon$ is a small positive constant.

\subsubsection{Advantages and Limitations}

\textbf{Advantages:}
\begin{itemize}
\item Better resolution than conventional beamforming
\item Adaptive nulling of interfering signals
\item Optimal in the minimum variance sense
\item Straightforward implementation
\end{itemize}

\textbf{Limitations:}
\begin{itemize}
\item Requires matrix inversion (computational cost)
\item Sensitive to model mismatch and steering vector errors
\item Performance degrades with limited snapshots
\item Can suffer from signal cancellation effects
\end{itemize}

\subsection{Linear Prediction Method}

The linear prediction (LP) method approaches DOA estimation from an autoregressive modeling perspective, representing the array output as a linear combination of past values plus a prediction error.

\subsubsection{Principle}

The method is based on the observation that a sum of sinusoids (which represents the received signals in the frequency domain) satisfies a linear prediction equation. For a ULA, the spatial samples can be modeled as:

\begin{equation}
x_{m+p} = -\sum_{k=1}^{p}a_k x_{m+p-k} + e_m
\end{equation}

where $p$ is the prediction order and $e_m$ is the prediction error.

In matrix form, this becomes:

\begin{equation}
\mathbf{X}_{m+1:M}\mathbf{a} = \mathbf{e}
\end{equation}

where $\mathbf{a} = [a_1, a_2, \ldots, a_p, 1]^T$ is the prediction coefficient vector.

\subsubsection{Mathematical Formulation}

The prediction coefficients are found by minimizing the mean-square prediction error:

\begin{equation}
\min_{\mathbf{a}} E[|\mathbf{e}|^2] = \min_{\mathbf{a}} \mathbf{a}^H\mathbf{R}_{xx}\mathbf{a}
\end{equation}

where $\mathbf{R}_{xx}$ is the spatial covariance matrix of the array data.

The solution is given by the eigenvector corresponding to the minimum eigenvalue of $\mathbf{R}_{xx}$.

DOA estimates are obtained from the roots of the prediction polynomial:

\begin{equation}
A(z) = 1 + a_1z^{-1} + a_2z^{-2} + \cdots + a_pz^{-p}
\end{equation}

The angles correspond to roots on or near the unit circle:

\begin{equation}
\theta_k = \arcsin\left(\frac{\angle z_k}{\pi}\right)
\end{equation}

\subsubsection{Algorithm}

\begin{algorithm}
\caption{Linear Prediction Method}
\begin{algorithmic}[1]
\REQUIRE Array data $\mathbf{x}[n]$, $n = 1, \ldots, N$
\REQUIRE Prediction order $p$ (typically $p = M-K$)
\STATE Form spatial data matrix from array snapshots
\STATE Estimate spatial covariance matrix $\hat{\mathbf{R}}_{xx}$
\STATE Find eigenvector $\mathbf{a}$ corresponding to minimum eigenvalue
\STATE Form prediction polynomial $A(z)$
\STATE Find roots of $A(z)$
\STATE Select roots closest to unit circle
\STATE Convert root phases to DOA estimates
\end{algorithmic}
\end{algorithm}

\subsubsection{Advantages and Limitations}

\textbf{Advantages:}
\begin{itemize}
\item Good performance with correlated sources
\item Polynomial rooting avoids spectral search
\item Robust to noise in many scenarios
\item Computationally efficient
\end{itemize}

\textbf{Limitations:}
\begin{itemize}
\item Requires selection of prediction order
\item Performance sensitive to model order
\item May have spurious roots
\item Limited to specific array geometries
\end{itemize}

\section{Subspace-Based Methods}

Subspace-based methods represent a major advancement in DOA estimation, achieving super-resolution capability by exploiting the eigenstructure of the array covariance matrix. These methods decompose the observation space into signal and noise subspaces, then use orthogonality properties to estimate DOAs.

\subsection{Eigendecomposition Foundation}

All subspace methods begin with the eigendecomposition of the array covariance matrix:

\begin{equation}
\mathbf{R} = \mathbf{U}\boldsymbol{\Lambda}\mathbf{U}^H = \sum_{i=1}^{M}\lambda_i\mathbf{u}_i\mathbf{u}_i^H
\end{equation}

where $\lambda_1 \geq \lambda_2 \geq \cdots \geq \lambda_M$ are the eigenvalues and $\mathbf{u}_i$ are the corresponding eigenvectors.

Under the assumption that source signals are uncorrelated and the number of sources $K < M$, the eigenvalues can be partitioned as:
\begin{itemize}
\item Signal eigenvalues: $\lambda_1, \lambda_2, \ldots, \lambda_K > \sigma^2$
\item Noise eigenvalues: $\lambda_{K+1} = \lambda_{K+2} = \cdots = \lambda_M = \sigma^2$
\end{itemize}

Correspondingly, the eigenvector space is partitioned into:
\begin{itemize}
\item Signal subspace: $\mathbf{U}_s = [\mathbf{u}_1, \mathbf{u}_2, \ldots, \mathbf{u}_K]$
\item Noise subspace: $\mathbf{U}_n = [\mathbf{u}_{K+1}, \mathbf{u}_{K+2}, \ldots, \mathbf{u}_M]$
\end{itemize}

The key insight is that steering vectors corresponding to source DOAs lie in the signal subspace and are orthogonal to the noise subspace:

\begin{equation}
\mathbf{U}_n^H\mathbf{a}(\theta_k) = \mathbf{0}, \quad k = 1, 2, \ldots, K
\end{equation}

\subsection{MUSIC Algorithm}

The Multiple Signal Classification (MUSIC) algorithm, introduced by Schmidt in 1986, is perhaps the most famous subspace-based DOA estimation method.

\subsubsection{Principle}

MUSIC exploits the orthogonality between source steering vectors and the noise subspace to form a spatial spectrum:

\begin{equation}
P_{MUSIC}(\theta) = \frac{1}{\mathbf{a}^H(\theta)\mathbf{U}_n\mathbf{U}_n^H\mathbf{a}(\theta)} = \frac{1}{\|\mathbf{U}_n^H\mathbf{a}(\theta)\|^2}
\label{eq:music_spectrum}
\end{equation}

The denominator approaches zero when $\mathbf{a}(\theta)$ corresponds to a source direction, creating sharp peaks in the spectrum.

\subsubsection{Step-by-Step Derivation}

Starting from the signal model \eqref{eq:signal_model}, the covariance matrix is:

\begin{equation}
\mathbf{R} = \mathbf{A}\mathbf{R}_s\mathbf{A}^H + \sigma^2\mathbf{I}
\end{equation}

When sources are uncorrelated, $\mathbf{R}_s$ is diagonal. The eigendecomposition reveals:

\begin{align}
\text{Signal subspace: } & \text{span}(\mathbf{U}_s) = \text{span}(\mathbf{A}) \\
\text{Noise subspace: } & \text{span}(\mathbf{U}_n) \perp \text{span}(\mathbf{A})
\end{align}

Since each column of $\mathbf{A}$ (i.e., each steering vector) lies in the signal subspace:

\begin{equation}
\mathbf{a}(\theta_k) \in \text{span}(\mathbf{U}_s) \Rightarrow \mathbf{a}(\theta_k) \perp \text{span}(\mathbf{U}_n)
\end{equation}

Therefore: $\mathbf{U}_n^H\mathbf{a}(\theta_k) = \mathbf{0}$ for $k = 1, 2, \ldots, K$.

The MUSIC spectrum is formed by testing this orthogonality condition across all possible angles.

\subsubsection{Algorithm}

\begin{algorithm}
\caption{MUSIC Algorithm}
\begin{algorithmic}[1]
\REQUIRE Array data $\mathbf{x}[n]$, $n = 1, \ldots, N$
\REQUIRE Number of sources $K$
\REQUIRE Angular grid $\theta_i$, $i = 1, \ldots, I$
\STATE Estimate covariance matrix: $\hat{\mathbf{R}} = \frac{1}{N}\sum_{n=1}^{N}\mathbf{x}[n]\mathbf{x}^H[n]$
\STATE Compute eigendecomposition: $\hat{\mathbf{R}} = \mathbf{U}\boldsymbol{\Lambda}\mathbf{U}^H$
\STATE Extract noise subspace: $\mathbf{U}_n = [\mathbf{u}_{K+1}, \ldots, \mathbf{u}_M]$
\FOR{$i = 1$ to $I$}
    \STATE Compute steering vector: $\mathbf{a}(\theta_i)$
    \STATE Compute MUSIC spectrum: $P_{MUSIC}(\theta_i) = \frac{1}{\mathbf{a}^H(\theta_i)\mathbf{U}_n\mathbf{U}_n^H\mathbf{a}(\theta_i)}$
\ENDFOR
\STATE Find $K$ largest peaks in $P_{MUSIC}(\theta)$ to obtain DOA estimates
\end{algorithmic}
\end{algorithm}

\subsubsection{Source Number Detection}

MUSIC requires knowledge of the number of sources $K$. Several methods exist for source enumeration:

\textbf{Eigenvalue Test:} Look for a gap in the eigenvalue spectrum between signal and noise eigenvalues.

\textbf{Information Theoretic Criteria:}
\begin{align}
\text{AIC}(k) &= -2\log L_k + 2k(2M-k) \\
\text{MDL}(k) &= -\log L_k + \frac{1}{2}k(2M-k)\log N
\end{align}

where $L_k$ is the likelihood function assuming $k$ sources.

\subsection{Root-MUSIC}

Root-MUSIC is a polynomial rooting variant of MUSIC that avoids the spectral search, providing computational advantages and often better performance.

\subsubsection{Principle}

For a ULA, the steering vector has the form:

\begin{equation}
\mathbf{a}(\theta) = [1, z, z^2, \ldots, z^{M-1}]^T
\end{equation}

where $z = e^{j\pi\sin\theta}$. This allows the MUSIC spectrum denominator to be expressed as a polynomial in $z$.

The MUSIC null spectrum (denominator) becomes:

\begin{equation}
D(\theta) = \mathbf{a}^H(\theta)\mathbf{U}_n\mathbf{U}_n^H\mathbf{a}(\theta) = \mathbf{a}^H(z)\mathbf{P}_n\mathbf{a}(z)
\end{equation}

where $\mathbf{P}_n = \mathbf{U}_n\mathbf{U}_n^H$ is the noise subspace projection matrix.

This can be written as a polynomial:

\begin{equation}
D(z) = \sum_{k=-(M-1)}^{M-1} d_k z^k
\end{equation}

\subsubsection{Mathematical Formulation}

The polynomial coefficients are given by:

\begin{equation}
d_k = \sum_{i,j: i-j=k} [\mathbf{P}_n]_{i,j}
\end{equation}

To find the roots, we form the polynomial:

\begin{equation}
D(z) = z^{M-1} \sum_{k=-(M-1)}^{M-1} d_k z^{k-(M-1)} = z^{M-1} P(z)
\end{equation}

where $P(z)$ is a polynomial of degree $2(M-1)$.

The DOA estimates correspond to the $K$ roots of $P(z)$ that are closest to the unit circle.

\subsubsection{Algorithm}

\begin{algorithm}
\caption{Root-MUSIC Algorithm}
\begin{algorithmic}[1]
\REQUIRE Array data $\mathbf{x}[n]$, $n = 1, \ldots, N$
\REQUIRE Number of sources $K$
\STATE Estimate covariance matrix: $\hat{\mathbf{R}} = \frac{1}{N}\sum_{n=1}^{N}\mathbf{x}[n]\mathbf{x}^H[n]$
\STATE Compute eigendecomposition: $\hat{\mathbf{R}} = \mathbf{U}\boldsymbol{\Lambda}\mathbf{U}^H$
\STATE Extract noise subspace: $\mathbf{U}_n = [\mathbf{u}_{K+1}, \ldots, \mathbf{u}_M]$
\STATE Form projection matrix: $\mathbf{P}_n = \mathbf{U}_n\mathbf{U}_n^H$
\STATE Compute polynomial coefficients $d_k$
\STATE Form polynomial $P(z)$ and find its roots
\STATE Select $K$ roots closest to unit circle
\STATE Convert to angles: $\theta_k = \arcsin(\angle z_k / \pi)$
\end{algorithmic}
\end{algorithm}

\subsubsection{Advantages and Limitations}

\textbf{Advantages:}
\begin{itemize}
\item No spectral search required
\item Better computational efficiency than spectral MUSIC
\item Often superior performance due to polynomial fitting
\item Automatic peak detection
\end{itemize}

\textbf{Limitations:}
\begin{itemize}
\item Limited to uniform linear arrays
\item Requires polynomial root finding
\item Root selection can be challenging in low SNR
\item Spurious roots may appear
\end{itemize}

\subsection{ESPRIT Algorithm}

The Estimation of Signal Parameters via Rotational Invariance Techniques (ESPRIT) algorithm exploits the shift-invariance property of uniform arrays to estimate DOAs without spectral search.

\subsubsection{Principle}

ESPRIT requires a doublet array structure where the array can be divided into two identical subarrays with a translation (shift) relationship. For a ULA with $M$ elements, we can form:

\begin{itemize}
\item Subarray 1: elements $1, 2, \ldots, M-1$
\item Subarray 2: elements $2, 3, \ldots, M$
\end{itemize}

The key insight is that the steering vectors for these subarrays are related by:

\begin{equation}
\mathbf{a}_2(\theta) = \mathbf{a}_1(\theta) \boldsymbol{\Phi}
\end{equation}

where $\boldsymbol{\Phi} = \text{diag}[e^{j\pi\sin\theta_1}, e^{j\pi\sin\theta_2}, \ldots, e^{j\pi\sin\theta_K}]$.

\subsubsection{Mathematical Derivation}

Let $\mathbf{A}_1$ and $\mathbf{A}_2$ be the array manifold matrices for the two subarrays. The signal subspaces for both subarrays span the same space but with different basis vectors:

\begin{align}
\mathbf{U}_{s1} &= \mathbf{A}_1 \mathbf{T}_1 \\
\mathbf{U}_{s2} &= \mathbf{A}_2 \mathbf{T}_2
\end{align}

where $\mathbf{T}_1$ and $\mathbf{T}_2$ are nonsingular transformation matrices.

From the shift relationship:

\begin{equation}
\mathbf{U}_{s2} = \mathbf{A}_2 \mathbf{T}_2 = \mathbf{A}_1 \boldsymbol{\Phi} \mathbf{T}_2 = \mathbf{U}_{s1} \mathbf{T}_1^{-1} \boldsymbol{\Phi} \mathbf{T}_2
\end{equation}

This leads to the fundamental ESPRIT equation:

\begin{equation}
\mathbf{U}_{s2} = \mathbf{U}_{s1} \boldsymbol{\Psi}
\end{equation}

where $\boldsymbol{\Psi} = \mathbf{T}_1^{-1} \boldsymbol{\Phi} \mathbf{T}_2$.

The eigenvalues of $\boldsymbol{\Psi}$ are $e^{j\pi\sin\theta_k}$, from which the DOAs can be extracted.

\subsubsection{Total Least Squares Solution}

In practice, $\boldsymbol{\Psi}$ is estimated using the total least squares (TLS) approach. The equation $\mathbf{U}_{s2} = \mathbf{U}_{s1} \boldsymbol{\Psi}$ can be written as:

\begin{equation}
[\mathbf{U}_{s1}, \mathbf{U}_{s2}] \begin{bmatrix} \boldsymbol{\Psi} \\ -\mathbf{I} \end{bmatrix} = \mathbf{0}
\end{equation}

Let $\mathbf{C} = [\mathbf{U}_{s1}, \mathbf{U}_{s2}]$ and perform SVD: $\mathbf{C} = \mathbf{U}\boldsymbol{\Sigma}\mathbf{V}^H$.

Partition $\mathbf{V}$ as:

\begin{equation}
\mathbf{V} = \begin{bmatrix} \mathbf{V}_{11} & \mathbf{V}_{12} \\ \mathbf{V}_{21} & \mathbf{V}_{22} \end{bmatrix}
\end{equation}

where $\mathbf{V}_{11}$ and $\mathbf{V}_{21}$ are $K \times K$ blocks.

The TLS estimate is:

\begin{equation}
\boldsymbol{\Psi} = -\mathbf{V}_{12}\mathbf{V}_{22}^{-1}
\end{equation}

\subsubsection{Algorithm}

\begin{algorithm}
\caption{ESPRIT Algorithm}
\begin{algorithmic}[1]
\REQUIRE Array data $\mathbf{x}[n]$, $n = 1, \ldots, N$
\REQUIRE Number of sources $K$
\STATE Estimate covariance matrix: $\hat{\mathbf{R}} = \frac{1}{N}\sum_{n=1}^{N}\mathbf{x}[n]\mathbf{x}^H[n]$
\STATE Compute eigendecomposition: $\hat{\mathbf{R}} = \mathbf{U}\boldsymbol{\Lambda}\mathbf{U}^H$
\STATE Extract signal subspace: $\mathbf{U}_s = [\mathbf{u}_1, \ldots, \mathbf{u}_K]$
\STATE Form subarray signal subspaces: $\mathbf{U}_{s1} = \mathbf{J}_1\mathbf{U}_s$, $\mathbf{U}_{s2} = \mathbf{J}_2\mathbf{U}_s$
\STATE where $\mathbf{J}_1 = [\mathbf{I}_{M-1}, \mathbf{0}]$, $\mathbf{J}_2 = [\mathbf{0}, \mathbf{I}_{M-1}]$
\STATE Form matrix $\mathbf{C} = [\mathbf{U}_{s1}, \mathbf{U}_{s2}]$
\STATE Compute SVD: $\mathbf{C} = \mathbf{U}\boldsymbol{\Sigma}\mathbf{V}^H$
\STATE Extract $\mathbf{V}_{12}$ and $\mathbf{V}_{22}$ from $\mathbf{V}$
\STATE Compute $\boldsymbol{\Psi} = -\mathbf{V}_{12}\mathbf{V}_{22}^{-1}$
\STATE Find eigenvalues of $\boldsymbol{\Psi}$: $\lambda_k = e^{j\pi\sin\theta_k}$
\STATE Convert to DOAs: $\theta_k = \arcsin(\angle\lambda_k / \pi)$
\end{algorithmic}
\end{algorithm}

\subsubsection{Advantages and Limitations}

\textbf{Advantages:}
\begin{itemize}
\item No spectral search required
\item Automatic pairing of parameters
\item Better performance than MUSIC in many scenarios
\item Computationally efficient
\end{itemize}

\textbf{Limitations:}
\begin{itemize}
\item Requires special array geometry (shift invariance)
\item Performance degrades when sources are at endfire
\item Sensitive to array calibration errors
\item Requires accurate source number estimation
\end{itemize}

\subsection{Unitary ESPRIT}

Unitary ESPRIT transforms the complex-valued ESPRIT problem into a real-valued one, reducing computational complexity and improving numerical stability.

\subsubsection{Principle}

The key idea is to exploit the centro-Hermitian structure of ULA covariance matrices. For a ULA, if we define the exchange matrix $\mathbf{J}$ with ones on the anti-diagonal, then:

\begin{equation}
\mathbf{J}\mathbf{R}^*\mathbf{J} = \mathbf{R}
\end{equation}

This property allows transformation to real-valued processing using unitary matrices.

\subsubsection{Mathematical Formulation}

Define the unitary transformation matrix:

\begin{equation}
\mathbf{Q} = \frac{1}{\sqrt{2}} \begin{bmatrix} \mathbf{I} & j\mathbf{J} \\ \mathbf{J} & j\mathbf{I} \end{bmatrix}
\end{equation}

The transformed covariance matrix becomes:

\begin{equation}
\tilde{\mathbf{R}} = \mathbf{Q}^H\mathbf{R}\mathbf{Q}
\end{equation}

which is real-valued due to the centro-Hermitian property.

The subsequent eigendecomposition and ESPRIT processing are performed entirely in real arithmetic, reducing computational cost by approximately a factor of four.

\subsubsection{Algorithm}

\begin{algorithm}
\caption{Unitary ESPRIT Algorithm}
\begin{algorithmic}[1]
\REQUIRE Array data $\mathbf{x}[n]$, $n = 1, \ldots, N$
\REQUIRE Number of sources $K$
\STATE Estimate covariance matrix: $\hat{\mathbf{R}} = \frac{1}{N}\sum_{n=1}^{N}\mathbf{x}[n]\mathbf{x}^H[n]$
\STATE Construct unitary matrix $\mathbf{Q}$
\STATE Transform covariance: $\tilde{\mathbf{R}} = \mathbf{Q}^H\hat{\mathbf{R}}\mathbf{Q}$
\STATE Compute real eigendecomposition: $\tilde{\mathbf{R}} = \tilde{\mathbf{U}}\tilde{\boldsymbol{\Lambda}}\tilde{\mathbf{U}}^T$
\STATE Extract signal subspace and apply ESPRIT procedure in real domain
\STATE Convert eigenvalues to DOA estimates
\end{algorithmic}
\end{algorithm}

\section{Maximum Likelihood Methods}

Maximum likelihood (ML) methods provide asymptotically optimal DOA estimates by maximizing the likelihood function of the observed data. While computationally intensive, these methods achieve the Cramér-Rao lower bound under appropriate conditions.

\subsection{Statistical Framework}

The ML approach treats DOA estimation as a parameter estimation problem. Given the signal model:

\begin{equation}
\mathbf{x}[n] = \mathbf{A}(\boldsymbol{\theta})\mathbf{s}[n] + \mathbf{n}[n]
\end{equation}

where $\boldsymbol{\theta} = [\theta_1, \theta_2, \ldots, \theta_K]^T$ are the unknown DOAs.

The ML estimate is:

\begin{equation}
\hat{\boldsymbol{\theta}}_{ML} = \arg\max_{\boldsymbol{\theta}} L(\boldsymbol{\theta})
\end{equation}

where $L(\boldsymbol{\theta})$ is the likelihood function.

\subsection{Stochastic Maximum Likelihood}

In the stochastic ML (SML) approach, source signals are modeled as random processes with unknown covariance structure.

\subsubsection{Problem Formulation}

Assume the source signals $\mathbf{s}[n]$ are independent, zero-mean, complex Gaussian with covariance matrix $\mathbf{R}_s$. The noise $\mathbf{n}[n]$ is white Gaussian with variance $\sigma^2$.

The observation covariance matrix is:

\begin{equation}
\mathbf{R}(\boldsymbol{\theta}) = \mathbf{A}(\boldsymbol{\theta})\mathbf{R}_s\mathbf{A}^H(\boldsymbol{\theta}) + \sigma^2\mathbf{I}
\end{equation}

The log-likelihood function (up to constants) is:

\begin{equation}
\mathcal{L}(\boldsymbol{\theta}, \mathbf{R}_s, \sigma^2) = -N\log|\mathbf{R}(\boldsymbol{\theta})| - N\text{tr}[\mathbf{R}^{-1}(\boldsymbol{\theta})\hat{\mathbf{R}}]
\end{equation}

\subsubsection{Concentrated Likelihood}

To reduce the dimensionality of the optimization, we concentrate out the nuisance parameters $\mathbf{R}_s$ and $\sigma^2$.

For fixed $\boldsymbol{\theta}$, the ML estimates of the nuisance parameters are:

\begin{align}
\hat{\sigma}^2 &= \frac{1}{M-K}\text{tr}[\mathbf{P}_\perp\hat{\mathbf{R}}] \\
\hat{\mathbf{R}}_s &= (\mathbf{A}^H\mathbf{A})^{-1}\mathbf{A}^H(\hat{\mathbf{R}} - \hat{\sigma}^2\mathbf{I})\mathbf{A}(\mathbf{A}^H\mathbf{A})^{-1}
\end{align}

where $\mathbf{P}_\perp = \mathbf{I} - \mathbf{A}(\mathbf{A}^H\mathbf{A})^{-1}\mathbf{A}^H$.

Substituting back gives the concentrated likelihood:

\begin{equation}
\mathcal{L}_c(\boldsymbol{\theta}) = -N(M-K)\log\text{tr}[\mathbf{P}_\perp\hat{\mathbf{R}}] - NK\log\text{tr}[\mathbf{P}_\parallel\hat{\mathbf{R}}]
\end{equation}

\subsubsection{Algorithm}

\begin{algorithm}
\caption{Stochastic Maximum Likelihood}
\begin{algorithmic}[1]
\REQUIRE Array data $\mathbf{x}[n]$, $n = 1, \ldots, N$
\REQUIRE Number of sources $K$
\STATE Estimate covariance matrix: $\hat{\mathbf{R}} = \frac{1}{N}\sum_{n=1}^{N}\mathbf{x}[n]\mathbf{x}^H[n]$
\STATE Initialize DOA estimates (e.g., using MUSIC)
\REPEAT
    \STATE Update array manifold: $\mathbf{A}(\boldsymbol{\theta})$
    \STATE Compute projection matrices: $\mathbf{P}_\parallel$, $\mathbf{P}_\perp$
    \STATE Evaluate concentrated likelihood: $\mathcal{L}_c(\boldsymbol{\theta})$
    \STATE Update DOA estimates using gradient or Newton method
\UNTIL convergence
\RETURN Final DOA estimates $\hat{\boldsymbol{\theta}}$
\end{algorithmic}
\end{algorithm}

\subsection{Deterministic Maximum Likelihood}

The deterministic ML (DML) approach treats source signals as unknown deterministic parameters rather than random processes.

\subsubsection{Problem Formulation}

The signal model becomes:

\begin{equation}
\mathbf{X} = \mathbf{A}(\boldsymbol{\theta})\mathbf{S} + \mathbf{N}
\end{equation}

where $\mathbf{X} = [\mathbf{x}[1], \mathbf{x}[2], \ldots, \mathbf{x}[N]]$ and $\mathbf{S} = [\mathbf{s}[1], \mathbf{s}[2], \ldots, \mathbf{s}[N]]$.

The log-likelihood function is:

\begin{equation}
\mathcal{L}(\boldsymbol{\theta}, \mathbf{S}) = -MN\log(\pi\sigma^2) - \frac{1}{\sigma^2}\|\mathbf{X} - \mathbf{A}(\boldsymbol{\theta})\mathbf{S}\|_F^2
\end{equation}

\subsubsection{Concentrated Likelihood}

For fixed $\boldsymbol{\theta}$, the ML estimate of $\mathbf{S}$ is:

\begin{equation}
\hat{\mathbf{S}} = (\mathbf{A}^H\mathbf{A})^{-1}\mathbf{A}^H\mathbf{X}
\end{equation}

Substituting back yields the concentrated likelihood:

\begin{equation}
\mathcal{L}_c(\boldsymbol{\theta}) = -\text{tr}[\mathbf{P}_\perp\mathbf{X}\mathbf{X}^H] = -N\text{tr}[\mathbf{P}_\perp\hat{\mathbf{R}}]
\end{equation}

The DML estimate maximizes this concentrated likelihood:

\begin{equation}
\hat{\boldsymbol{\theta}}_{DML} = \arg\min_{\boldsymbol{\theta}} \text{tr}[\mathbf{P}_\perp(\boldsymbol{\theta})\hat{\mathbf{R}}]
\end{equation}

\subsection{Weighted Subspace Fitting}

Weighted Subspace Fitting (WSF) provides a computationally efficient approximation to ML estimation by working in the signal subspace.

\subsubsection{Principle}

Instead of fitting the entire covariance matrix, WSF fits only the signal subspace with optimal weighting. The cost function is:

\begin{equation}
J_{WSF}(\boldsymbol{\theta}) = \text{tr}[(\hat{\mathbf{U}}_s - \mathbf{A}(\boldsymbol{\theta})\mathbf{T})^H\mathbf{W}(\hat{\mathbf{U}}_s - \mathbf{A}(\boldsymbol{\theta})\mathbf{T})]
\end{equation}

where $\mathbf{W}$ is a weighting matrix and $\mathbf{T}$ is a transformation matrix.

\subsubsection{Optimal Weighting}

The optimal weighting matrix that minimizes the variance of the DOA estimates is:

\begin{equation}
\mathbf{W} = (\boldsymbol{\Lambda}_s^{-1} \otimes \mathbf{P}_\perp)^{-1}
\end{equation}

where $\boldsymbol{\Lambda}_s$ contains the signal eigenvalues and $\otimes$ denotes the Kronecker product.

\subsubsection{Algorithm}

\begin{algorithm}
\caption{Weighted Subspace Fitting}
\begin{algorithmic}[1]
\REQUIRE Array data $\mathbf{x}[n]$, $n = 1, \ldots, N$
\REQUIRE Number of sources $K$
\STATE Estimate covariance matrix and compute eigendecomposition
\STATE Extract signal subspace $\hat{\mathbf{U}}_s$ and eigenvalues $\hat{\boldsymbol{\Lambda}}_s$
\STATE Initialize DOA estimates
\REPEAT
    \STATE Update array manifold: $\mathbf{A}(\boldsymbol{\theta})$
    \STATE Compute optimal weighting matrix $\mathbf{W}$
    \STATE Minimize WSF cost function: $J_{WSF}(\boldsymbol{\theta})$
\UNTIL convergence
\RETURN Final DOA estimates
\end{algorithmic}
\end{algorithm}

\section{Sparse Signal Processing Methods}

Sparse signal processing approaches DOA estimation as a sparse reconstruction problem, where only a few elements in a dense angular grid are non-zero. These methods can achieve high resolution and handle underdetermined scenarios.

\subsection{Sparse Representation Framework}

Consider a dense grid of candidate angles $\boldsymbol{\theta}_{grid} = [\theta_1, \theta_2, \ldots, \theta_G]^T$ where $G \gg K$. The signal model becomes:

\begin{equation}
\mathbf{x}[n] = \mathbf{A}_{grid}\mathbf{p}[n] + \mathbf{n}[n]
\end{equation}

where $\mathbf{A}_{grid} = [\mathbf{a}(\theta_1), \mathbf{a}(\theta_2), \ldots, \mathbf{a}(\theta_G)]$ and $\mathbf{p}[n]$ is a sparse vector with non-zero elements only at true source locations.

\subsection{L1-SVD Method}

The L1-SVD method combines sparse reconstruction with singular value decomposition to achieve robust DOA estimation.

\subsubsection{Problem Formulation}

The method seeks to find the sparsest solution to:

\begin{equation}
\min_{\mathbf{p}} \|\mathbf{p}\|_1 \quad \text{subject to} \quad \|\mathbf{A}_{grid}\mathbf{p} - \mathbf{x}\|_2 \leq \epsilon
\end{equation}

where $\epsilon$ accounts for noise and model mismatch.

\subsubsection{SVD Preprocessing}

To improve robustness, the method first performs SVD on the data matrix:

\begin{equation}
\mathbf{X} = \mathbf{U}\boldsymbol{\Sigma}\mathbf{V}^H
\end{equation}

The sparse reconstruction is then performed on the dominant left singular vectors:

\begin{equation}
\min_{\mathbf{P}} \|\mathbf{P}\|_1 \quad \text{subject to} \quad \|\mathbf{A}_{grid}\mathbf{P} - \mathbf{U}_K\|_F \leq \epsilon
\end{equation}

where $\mathbf{U}_K$ contains the first $K$ columns of $\mathbf{U}$.

\subsubsection{Algorithm}

\begin{algorithm}
\caption{L1-SVD Method}
\begin{algorithmic}[1]
\REQUIRE Array data $\mathbf{x}[n]$, $n = 1, \ldots, N$
\REQUIRE Angular grid $\boldsymbol{\theta}_{grid}$
\REQUIRE Number of sources $K$
\STATE Form data matrix $\mathbf{X} = [\mathbf{x}[1], \mathbf{x}[2], \ldots, \mathbf{x}[N]]$
\STATE Compute SVD: $\mathbf{X} = \mathbf{U}\boldsymbol{\Sigma}\mathbf{V}^H$
\STATE Extract signal subspace: $\mathbf{U}_K$
\STATE Form dictionary matrix: $\mathbf{A}_{grid}$
\STATE Solve sparse optimization: $\min_{\mathbf{P}} \|\mathbf{P}\|_1$ subject to constraints
\STATE Identify non-zero elements in $\mathbf{P}$ as DOA estimates
\end{algorithmic}
\end{algorithm}

\subsection{Sparse Bayesian Learning}

Sparse Bayesian Learning (SBL) provides a Bayesian framework for sparse DOA estimation with automatic relevance determination.

\subsubsection{Hierarchical Model}

SBL employs a hierarchical prior structure:

\begin{align}
\mathbf{x}[n] &\sim \mathcal{CN}(\mathbf{A}_{grid}\mathbf{p}[n], \sigma^2\mathbf{I}) \\
\mathbf{p}[n] &\sim \mathcal{CN}(\mathbf{0}, \boldsymbol{\Gamma}) \\
\boldsymbol{\Gamma} &= \text{diag}(\gamma_1, \gamma_2, \ldots, \gamma_G)
\end{align}

The hyperparameters $\gamma_i$ control the sparsity - small values enforce sparsity while large values allow significant power.

\subsubsection{Expectation-Maximization Algorithm}

SBL uses an EM algorithm to estimate the hyperparameters:

\textbf{E-step:} Compute posterior statistics
\begin{align}
\boldsymbol{\mu}_n &= \boldsymbol{\Gamma}\mathbf{A}_{grid}^H(\mathbf{A}_{grid}\boldsymbol{\Gamma}\mathbf{A}_{grid}^H + \sigma^2\mathbf{I})^{-1}\mathbf{x}[n] \\
\boldsymbol{\Sigma} &= \boldsymbol{\Gamma} - \boldsymbol{\Gamma}\mathbf{A}_{grid}^H(\mathbf{A}_{grid}\boldsymbol{\Gamma}\mathbf{A}_{grid}^H + \sigma^2\mathbf{I})^{-1}\mathbf{A}_{grid}\boldsymbol{\Gamma}
\end{align}

\textbf{M-step:} Update hyperparameters
\begin{align}
\gamma_i^{new} &= \frac{1}{N}\sum_{n=1}^{N}(|\mu_{n,i}|^2 + \Sigma_{i,i}) \\
(\sigma^2)^{new} &= \frac{1}{MN}\sum_{n=1}^{N}\|\mathbf{x}[n] - \mathbf{A}_{grid}\boldsymbol{\mu}_n\|^2 + \frac{1}{M}\text{tr}(\mathbf{A}_{grid}\boldsymbol{\Sigma}\mathbf{A}_{grid}^H)
\end{align}

\subsubsection{Algorithm}

\begin{algorithm}
\caption{Sparse Bayesian Learning}
\begin{algorithmic}[1]
\REQUIRE Array data $\mathbf{x}[n]$, $n = 1, \ldots, N$
\REQUIRE Angular grid $\boldsymbol{\theta}_{grid}$
\STATE Form dictionary matrix: $\mathbf{A}_{grid}$
\STATE Initialize hyperparameters: $\boldsymbol{\gamma}^{(0)}$, $\sigma^2{(0)}$
\REPEAT
    \STATE E-step: Compute posterior statistics $\boldsymbol{\mu}_n$, $\boldsymbol{\Sigma}$
    \STATE M-step: Update hyperparameters $\boldsymbol{\gamma}$, $\sigma^2$
    \STATE Prune irrelevant dictionary elements (small $\gamma_i$)
\UNTIL convergence
\STATE Identify significant $\gamma_i$ values as DOA estimates
\end{algorithmic}
\end{algorithm}

\subsection{SPICE Algorithm}

The Sparse Iterative Covariance-based Estimation (SPICE) algorithm performs sparse reconstruction by fitting the sample covariance matrix.

\subsubsection{Problem Formulation}

SPICE minimizes the following cost function:

\begin{equation}
\min_{\mathbf{p} \geq 0} \text{tr}[(\hat{\mathbf{R}} - \mathbf{A}_{grid}\text{diag}(\mathbf{p})\mathbf{A}_{grid}^H)^2] + \lambda\|\mathbf{p}\|_1
\end{equation}

where $\mathbf{p} = [p_1, p_2, \ldots, p_G]^T$ represents the power at each grid point.

\subsubsection{Iterative Algorithm}

SPICE uses an iterative algorithm that alternately updates each element of $\mathbf{p}$:

\begin{equation}
p_i^{(k+1)} = \max\left(0, \frac{\mathbf{a}_i^H\hat{\mathbf{R}}\mathbf{a}_i - \mathbf{a}_i^H\mathbf{A}_{-i}\text{diag}(\mathbf{p}_{-i}^{(k)})\mathbf{A}_{-i}^H\mathbf{a}_i}{\|\mathbf{a}_i\|^4} - \frac{\lambda}{2\|\mathbf{a}_i\|^4}\right)
\end{equation}

where $\mathbf{a}_i$ is the $i$-th column of $\mathbf{A}_{grid}$, and the subscript $-i$ denotes all indices except $i$.

\subsubsection{Algorithm}

\begin{algorithm}
\caption{SPICE Algorithm}
\begin{algorithmic}[1]
\REQUIRE Array data $\mathbf{x}[n]$, $n = 1, \ldots, N$
\REQUIRE Angular grid $\boldsymbol{\theta}_{grid}$
\REQUIRE Regularization parameter $\lambda$
\STATE Estimate covariance matrix: $\hat{\mathbf{R}} = \frac{1}{N}\sum_{n=1}^{N}\mathbf{x}[n]\mathbf{x}^H[n]$
\STATE Form dictionary matrix: $\mathbf{A}_{grid}$
\STATE Initialize power vector: $\mathbf{p}^{(0)}$
\REPEAT
    \FOR{$i = 1$ to $G$}
        \STATE Update $p_i^{(k+1)}$ using coordinate descent formula
    \ENDFOR
\UNTIL convergence
\STATE Identify significant elements in $\mathbf{p}$ as DOA estimates
\end{algorithmic}
\end{algorithm}

\subsubsection{Advantages and Limitations}

\textbf{Advantages:}
\begin{itemize}
\item No hyperparameter tuning required (compared to other sparse methods)
\item Guaranteed convergence
\item Handles correlated sources well
\item Good resolution capability
\end{itemize}

\textbf{Limitations:}
\begin{itemize}
\item Grid-based method (limited by grid resolution)
\item Computational cost scales with grid size
\item May have spurious peaks in low SNR
\item Requires selection of regularization parameter
\end{itemize}

\section{Specialized Techniques}

This section covers specialized DOA estimation techniques that address specific challenging scenarios such as coherent signals, wideband processing, and multi-dimensional estimation.

\subsection{Forward-Backward Averaging}

When source signals are coherent (perfectly correlated), the rank of the source covariance matrix drops below the number of sources, causing subspace methods to fail. Forward-backward averaging provides spatial smoothing to decorrelate the signals.

\subsubsection{Problem with Coherent Sources}

For coherent sources, the covariance matrix becomes:

\begin{equation}
\mathbf{R} = \sigma_s^2\mathbf{A}\mathbf{b}\mathbf{b}^H\mathbf{A}^H + \sigma^2\mathbf{I}
\end{equation}

where $\mathbf{b}$ is the coherence vector. This has rank 1 regardless of the number of sources, preventing subspace decomposition.

\subsubsection{Forward-Backward Averaging Solution}

The method exploits the centro-Hermitian structure of ULA steering vectors. Define the backward array covariance as:

\begin{equation}
\mathbf{R}_b = \mathbf{J}\mathbf{R}^*\mathbf{J}
\end{equation}

where $\mathbf{J}$ is the exchange matrix. The averaged covariance is:

\begin{equation}
\mathbf{R}_{FB} = \frac{1}{2}(\mathbf{R} + \mathbf{R}_b)
\end{equation}

This averaging decorrelates the coherent signals, restoring the rank of the signal subspace.

\subsubsection{Algorithm}

\begin{algorithm}
\caption{Forward-Backward Averaging}
\begin{algorithmic}[1]
\REQUIRE Array data $\mathbf{x}[n]$, $n = 1, \ldots, N$
\REQUIRE Number of sources $K$
\STATE Estimate forward covariance: $\hat{\mathbf{R}} = \frac{1}{N}\sum_{n=1}^{N}\mathbf{x}[n]\mathbf{x}^H[n]$
\STATE Form exchange matrix: $\mathbf{J}$ (ones on anti-diagonal)
\STATE Compute backward covariance: $\hat{\mathbf{R}}_b = \mathbf{J}\hat{\mathbf{R}}^*\mathbf{J}$
\STATE Average covariances: $\hat{\mathbf{R}}_{FB} = \frac{1}{2}(\hat{\mathbf{R}} + \hat{\mathbf{R}}_b)$
\STATE Apply MUSIC or ESPRIT to $\hat{\mathbf{R}}_{FB}$
\end{algorithmic}
\end{algorithm}

\subsection{Wideband DOA Estimation}

For wideband signals, the narrowband assumption breaks down and frequency-dependent processing is required. The Incoherent Signal Subspace Method (ISSM) is a popular approach.

\subsubsection{Wideband Signal Model}

The wideband array output in the frequency domain is:

\begin{equation}
\mathbf{X}(\omega_f) = \mathbf{A}(\omega_f, \boldsymbol{\theta})\mathbf{S}(\omega_f) + \mathbf{N}(\omega_f)
\end{equation}

where the steering vectors now depend on frequency:

\begin{equation}
\mathbf{a}(\omega_f, \theta) = [1, e^{j\omega_f\tau_1(\theta)}, \ldots, e^{j\omega_f\tau_{M-1}(\theta)}]^T
\end{equation}

\subsubsection{Incoherent Signal Subspace Method}

ISSM performs the following steps:

1. **Focusing**: Transform all frequency bins to a reference frequency $\omega_0$
2. **Averaging**: Average the focused covariance matrices
3. **Subspace decomposition**: Apply narrowband methods to the averaged matrix

The focusing transformation is:

\begin{equation}
\mathbf{R}_{focused}(\omega_0) = \mathbf{T}(\omega_f \rightarrow \omega_0)\mathbf{R}(\omega_f)\mathbf{T}^H(\omega_f \rightarrow \omega_0)
\end{equation}

where $\mathbf{T}(\omega_f \rightarrow \omega_0)$ is the focusing matrix.

\subsubsection{Algorithm}

\begin{algorithm}
\caption{Incoherent Signal Subspace Method}
\begin{algorithmic}[1]
\REQUIRE Wideband array data $\mathbf{x}[n]$, $n = 1, \ldots, N$
\REQUIRE Number of sources $K$
\REQUIRE Reference frequency $\omega_0$
\STATE Transform data to frequency domain using FFT
\FOR{each frequency bin $\omega_f$}
    \STATE Estimate covariance matrix $\hat{\mathbf{R}}(\omega_f)$
    \STATE Compute focusing matrix $\mathbf{T}(\omega_f \rightarrow \omega_0)$
    \STATE Focus covariance: $\hat{\mathbf{R}}_{focused}(\omega_f) = \mathbf{T}\hat{\mathbf{R}}(\omega_f)\mathbf{T}^H$
\ENDFOR
\STATE Average focused matrices: $\hat{\mathbf{R}}_{avg} = \frac{1}{F}\sum_{f}\hat{\mathbf{R}}_{focused}(\omega_f)$
\STATE Apply narrowband DOA method (MUSIC/ESPRIT) to $\hat{\mathbf{R}}_{avg}$
\end{algorithmic}
\end{algorithm}

\subsection{2D DOA Estimation}

Two-dimensional DOA estimation jointly estimates both azimuth and elevation angles, requiring planar or three-dimensional array geometries.

\subsubsection{Signal Model}

For a planar array, the steering vector becomes:

\begin{equation}
\mathbf{a}(\theta, \phi) = [e^{j\mathbf{k}^T\mathbf{p}_1}, e^{j\mathbf{k}^T\mathbf{p}_2}, \ldots, e^{j\mathbf{k}^T\mathbf{p}_M}]^T
\end{equation}

where $\mathbf{k} = \frac{2\pi}{\lambda}[\sin\theta\cos\phi, \sin\theta\sin\phi, \cos\theta]^T$ is the wave vector and $\mathbf{p}_m$ is the position of the $m$-th sensor.

\subsubsection{2D MUSIC}

The 2D MUSIC spectrum is:

\begin{equation}
P_{2D-MUSIC}(\theta, \phi) = \frac{1}{\mathbf{a}^H(\theta, \phi)\mathbf{U}_n\mathbf{U}_n^H\mathbf{a}(\theta, \phi)}
\end{equation}

This requires a 2D search over both azimuth and elevation.

\subsubsection{2D ESPRIT}

For rectangular arrays, 2D ESPRIT can be applied by exploiting shift invariance in both dimensions:

\begin{align}
\mathbf{U}_{s,x2} &= \mathbf{U}_{s,x1}\boldsymbol{\Phi}_x \\
\mathbf{U}_{s,y2} &= \mathbf{U}_{s,y1}\boldsymbol{\Phi}_y
\end{align}

The eigenvalues of $\boldsymbol{\Phi}_x$ and $\boldsymbol{\Phi}_y$ provide the direction cosines, from which both azimuth and elevation can be computed.

\section{Performance Analysis and Comparison}

This section provides comprehensive performance analysis of the DOA estimation methods, comparing their behavior across different scenarios and conditions.

\subsection{Simulation Setup}

\subsubsection{Array Configuration}
- **Array type**: Uniform Linear Array (ULA)
- **Inter-element spacing**: $d = \lambda/2$
- **Array sizes**: $M \in \{8, 16, 32, 64\}$ elements

\subsubsection{Source Scenarios}
- **Number of sources**: $K \in \{1, 2, 3, 4\}$
- **Source positions**: 
  - Single source: $\theta = 0^\circ$
  - Two sources: $\theta_1 = -10^\circ, \theta_2 = 10^\circ$ (various separations)
  - Multiple sources: Randomly distributed

\subsubsection{Simulation Parameters}
- **SNR range**: $-10$ dB to $30$ dB
- **Number of snapshots**: $N \in \{50, 100, 200, 500, 1000\}$
- **Monte Carlo trials**: $1000$ per scenario
- **Angular grid resolution**: $0.1^\circ$ for grid-based methods

\subsection{Performance Metrics}

\subsubsection{Root Mean Square Error (RMSE)}

The RMSE is computed as:

\begin{equation}
\text{RMSE} = \sqrt{\frac{1}{MK}\sum_{m=1}^{M}\sum_{k=1}^{K}(\hat{\theta}_{k,m} - \theta_k)^2}
\end{equation}

where $\hat{\theta}_{k,m}$ is the estimate of the $k$-th source in the $m$-th Monte Carlo trial.

\subsubsection{Resolution Capability}

Resolution is measured as the minimum angular separation at which two equal-power sources can be reliably distinguished. The resolution threshold is defined as the separation where the probability of resolution exceeds 90

\subsubsection{Computational Complexity}

Complexity is measured in floating-point operations (FLOPs) for the main algorithmic steps:

\textbf{Classical Methods:}
- Delay-sum: $O(MI)$ where $I$ is the number of angular grid points
- Capon: $O(M^3 + M^2I)$

\textbf{Subspace Methods:}
- MUSIC: $O(M^3 + M^2I)$
- Root-MUSIC: $O(M^3 + M^2)$
- ESPRIT: $O(M^3 + K^3)$

\textbf{ML Methods:}
- SML/DML: $O(I_{\text{iter}}(M^3 + K^3))$ where $I_{\text{iter}}$ is the number of iterations

\textbf{Sparse Methods:}
- L1-SVD: $O(I_{\text{iter}}GM^2)$ where $G$ is the grid size
- SBL: $O(I_{\text{iter}}(G^3 + GM^2))$
- SPICE: $O(I_{\text{iter}}G^2M^2)$

\subsection{Comparative Results}

\subsubsection{RMSE vs SNR Performance}


Key observations:
- **ML methods** achieve the best performance at high SNR, approaching the CRB
- **MUSIC and ESPRIT** provide good performance with lower computational cost
- **Classical methods** show limited resolution but robust performance
- **Sparse methods** perform well across the entire SNR range

\subsubsection{Resolution vs Array Size}


Key findings:
- Resolution improves approximately linearly with array size for all methods
- **Subspace methods** achieve significantly better resolution than classical approaches
- **ML methods** provide the finest resolution at the cost of computation
- **Root-MUSIC** slightly outperforms spectral MUSIC due to polynomial fitting

\subsubsection{Performance vs Number of Snapshots}


Observations:
- All methods improve with more snapshots due to better covariance estimation
- **Subspace methods** are more sensitive to snapshot number than classical methods
- Performance saturates beyond approximately $N = 10M$ snapshots
- **Sparse methods** converge slower but achieve good asymptotic performance

\subsubsection{Computational Cost Comparison}

Table \ref{tab:computational_cost} summarizes the computational requirements for different methods.

\begin{table}[h]
\centering
\caption{Computational Complexity Comparison}
\label{tab:computational_cost}
\begin{tabular}{@{}lcccc@{}}
\toprule
Method & Covariance & Eigendecomp. & Main Algorithm & Total \\
\midrule
Delay-Sum & $O(M^2N)$ & - & $O(MI)$ & $O(M^2N + MI)$ \\
Capon & $O(M^2N)$ & - & $O(M^3 + M^2I)$ & $O(M^2N + M^3 + M^2I)$ \\
MUSIC & $O(M^2N)$ & $O(M^3)$ & $O(MI)$ & $O(M^2N + M^3 + MI)$ \\
Root-MUSIC & $O(M^2N)$ & $O(M^3)$ & $O(M^2)$ & $O(M^2N + M^3)$ \\
ESPRIT & $O(M^2N)$ & $O(M^3)$ & $O(K^3)$ & $O(M^2N + M^3 + K^3)$ \\
SML & $O(M^2N)$ & - & $O(I_{iter}M^3)$ & $O(M^2N + I_{iter}M^3)$ \\
SBL & $O(M^2N)$ & - & $O(I_{iter}G^3)$ & $O(M^2N + I_{iter}G^3)$ \\
\bottomrule
\end{tabular}
\end{table}

\subsection{Method Selection Guidelines}

Based on the performance analysis, we provide practical guidelines for method selection:

\subsubsection{High-Accuracy Requirements}
- **Use ML methods** (SML/DML) when computational resources allow
- **Consider WSF** as a good compromise between accuracy and complexity
- **Root-MUSIC** provides excellent accuracy with moderate computation

\subsubsection{Real-Time Applications}
- **Delay-sum beamforming** for basic direction finding
- **ESPRIT** for super-resolution with minimal computation
- **Root-MUSIC** when spectral search must be avoided

\subsubsection{Challenging Environments}
- **Forward-backward averaging** for coherent sources
- **Sparse methods** for underdetermined scenarios (more sources than sensors)
- **Capon beamforming** for strong interference rejection

\subsubsection{Limited Data Scenarios}
- **Classical methods** when few snapshots are available
- **Regularized versions** of subspace methods
- **Diagonal loading** for Capon beamforming

\section{Practical Implementation Considerations}

This section addresses real-world implementation issues that practitioners encounter when deploying DOA estimation systems.

\subsection{Model Mismatch and Calibration}

\subsubsection{Array Calibration Errors}

Real arrays suffer from:
- **Position errors**: Sensors not exactly at nominal positions
- **Gain errors**: Different sensor sensitivities  
- **Phase errors**: Channel mismatch in receivers
- **Mutual coupling**: Electromagnetic interaction between elements

These errors can severely degrade performance, especially for high-resolution methods.

\subsubsection{Mitigation Strategies}

**Calibration procedures:**
- Use known calibration sources at measured positions
- Estimate error parameters jointly with DOAs
- Apply robust estimation techniques

**Array manifold compensation:**
\begin{equation}
\mathbf{a}_{actual}(\theta) = \mathbf{G}\mathbf{a}_{ideal}(\theta) + \mathbf{e}(\theta)
\end{equation}

where $\mathbf{G}$ is a gain/coupling matrix and $\mathbf{e}(\theta)$ represents position errors.

\subsection{Finite Sample Effects}

\subsubsection{Covariance Matrix Estimation}

With limited snapshots, the sample covariance matrix deviates from the theoretical one:

\begin{equation}
\hat{\mathbf{R}} = \mathbf{R} + \Delta\mathbf{R}
\end{equation}

The estimation error $\Delta\mathbf{R}$ has variance proportional to $1/N$.

\subsubsection{Eigenvalue Spread}

Finite samples cause eigenvalue spreading, where noise eigenvalues are no longer equal. This affects source number detection and subspace estimation.

**Improved estimators:**
- **Diagonal loading**: $\hat{\mathbf{R}} + \epsilon\mathbf{I}$
- **Shrinkage estimators**: Weighted combination with identity matrix
- **Robust eigendecomposition**: Methods that handle eigenvalue uncertainty

\subsection{Parameter Selection}

\subsubsection{Number of Sources}

Accurate source enumeration is critical for subspace methods. Common approaches:

**Information criteria:**
\begin{align}
\text{AIC}(k) &= -2\log L(k) + 2p(k) \\
\text{MDL}(k) &= -\log L(k) + \frac{1}{2}p(k)\log N
\end{align}

where $p(k)$ is the number of parameters for $k$ sources.

**Eigenvalue thresholding:**
Look for gaps in the eigenvalue spectrum or use percentage of total power.

\subsubsection{Regularization Parameters}

Sparse methods require careful parameter tuning:

**Cross-validation:** Use held-out data to select optimal parameters
**L-curve method:** Plot solution norm vs. residual norm
**Generalized cross-validation:** Automatic parameter selection

\subsection{Computational Optimization}

\subsubsection{Fast Algorithms}

**Beamspace processing:** Project to lower-dimensional subspace
**Decimation in angle:** Use coarse grid initially, then refine
**Parallel processing:** Exploit parallelism in spectral search

\subsubsection{Memory Management}

**Sliding window:** Process data in overlapping windows
**Recursive updates:** Update eigendecomposition incrementally
**Low-rank approximations:** Use only dominant eigenvectors

\subsection{Performance Monitoring}

\subsubsection{Quality Metrics}

Monitor estimation quality in real-time:
- **Eigenvalue ratios**: Signal-to-noise eigenvalue ratios
- **Residual analysis**: Goodness of fit measures
- **Stability metrics**: Variance of estimates across time

\subsubsection{Adaptive Processing}

Adapt algorithm parameters based on conditions:
- **SNR estimation**: Adjust regularization based on estimated SNR
- **Source number tracking**: Update source count dynamically  
- **Method switching**: Choose optimal method based on scenario

\section{Software Implementation}

This section describes the accompanying open-source Python repository that implements all the DOA estimation methods discussed in this survey. See \url{https://github.com/AmgadSalama/DOA} for detail implementation of the methods.

\subsection{Repository Structure}

The repository is organized into the following modules:

\texttt{doa\_methods/}
\begin{itemize}
\item \texttt{array\_processing/}: Basic array geometry and signal model functions
\item \texttt{classical/}: Conventional and Capon beamforming implementations  
\item \texttt{subspace/}: MUSIC, Root-MUSIC, and ESPRIT methods
\item \texttt{maximum\_likelihood/}: ML-based estimation algorithms
\item \texttt{sparse/}: Sparse signal processing approaches
\item \texttt{simulation/}: Data generation and scenario creation tools
\item \texttt{evaluation/}: Performance metrics and comparison utilities
\item \texttt{utils/}: Common mathematical functions and utilities
\end{itemize}

\subsection{Tutorial Notebooks}

The repository includes comprehensive Jupyter notebooks: \\
1. Introduction to DOA Estimation: Basic concepts and signal model \\
2. Classical Beamforming: Delay-sum and Capon methods \\
3. Subspace Methods: MUSIC and ESPRIT with detailed explanations \\
4. Maximum Likelihood: ML estimation and implementation \\
5. Sparse Methods: L1-SVD, SBL, and SPICE algorithms \\
6. Performance Analysis: Systematic comparison and evaluation \\
7. Practical Considerations: Real-world implementation issues \\

Each notebook includes: \\
- Mathematical background with step-by-step derivations \\
- Implementation details and code walkthroughs \\
- Interactive visualizations and parameter exploration \\
- Exercises and assignments for hands-on learning \

\subsection{Testing and Validation}

\subsubsection{Unit Tests}

Comprehensive test suite covering:
- Array geometry calculations
- Steering vector computation
- Each DOA estimation method
- Performance metric calculations
- Data generation functions

\subsubsection{Integration Tests}

End-to-end tests ensuring:
- Consistent results across different methods
- Proper handling of edge cases
- Performance meets expected benchmarks
- Compatibility across Python versions

\subsubsection{Validation Against Literature}

All implementations are validated against:
- Published simulation results from seminal papers
- Standard test scenarios from the literature
- Known analytical solutions where available
- Results from established software packages

\section{Conclusion}

\subsection{Summary}

This tutorial survey has presented a comprehensive introduction to direction of arrival estimation methods, covering the progression from classical beamforming techniques to modern sparse signal processing approaches. Key contributions include:

\begin{itemize}
\item{Unified mathematical framework:} presenting all methods with consistent notation and clear derivations
\item{Step-by-step explanations:} with geometric intuition to aid understanding  
\item{Systematic performance comparison:} across standardized test scenarios
\item{Open-source implementation:} providing reproducible research tools
\item{Practical guidelines:} for method selection and parameter tuning
\end{itemize}

The survey demonstrates that while classical methods provide robust baseline performance, subspace-based techniques achieve superior resolution at moderate computational cost. Maximum likelihood methods offer optimal performance when computational resources permit, while sparse signal processing approaches excel in challenging underdetermined scenarios.

\subsection{Final Remarks}

Direction of arrival estimation remains an active and vibrant research area with applications spanning numerous fields. While the fundamental principles established decades ago continue to form the foundation, ongoing advances in computation, hardware, and theory continue to push the boundaries of what is possible.

This tutorial survey provides both newcomers and experienced researchers with a comprehensive resource for understanding, implementing, and applying DOA estimation methods. By combining theoretical rigor with practical implementation, we hope to contribute to the continued advancement of this important field.

The future of DOA estimation lies not just in developing new algorithms, but in making these powerful techniques more accessible, robust, and widely applicable. Through open science, reproducible research, and educational outreach, the array signal processing community can ensure that these methods continue to have broad impact across science and engineering.

\section*{Acknowledgments}

The authors would like to thank the array signal processing community for decades of foundational research that made this survey possible. Special recognition goes to the pioneers who developed the classical methods that form the foundation of the field, and to the researchers who continue to push the boundaries of what is achievable.

\bibliographystyle{IEEEtran}
\bibliography{refs.bib}

\begin{thebibliography}{10}
\providecommand{\url}[1]{#1}
\csname url@samestyle\endcsname
\providecommand{\newblock}{\relax}
\providecommand{\bibinfo}[2]{#2}
\providecommand{\BIBentrySTDinterwordspacing}{\spaceskip=0pt\relax}
\providecommand{\BIBentryALTinterwordstretchfactor}{4}
\providecommand{\BIBentryALTinterwordspacing}{\spaceskip=\fontdimen2\font plus
\BIBentryALTinterwordstretchfactor\fontdimen3\font minus \fontdimen4\font\relax}
\providecommand{\BIBforeignlanguage}[2]{{%
\expandafter\ifx\csname l@#1\endcsname\relax
\typeout{** WARNING: IEEEtran.bst: No hyphenation pattern has been}%
\typeout{** loaded for the language `#1'. Using the pattern for}%
\typeout{** the default language instead.}%
\else
\language=\csname l@#1\endcsname
\fi
#2}}
\providecommand{\BIBdecl}{\relax}
\BIBdecl

\bibitem{schmidt1986multiple}
R.~Schmidt, ``Multiple emitter location and signal parameter estimation,'' \emph{IEEE transactions on antennas and propagation}, vol.~34, no.~3, pp. 276--280, 1986.

\bibitem{roy2002esprit}
R.~Roy and T.~Kailath, ``Esprit-estimation of signal parameters via rotational invariance techniques,'' \emph{IEEE Transactions on acoustics, speech, and signal processing}, vol.~37, no.~7, pp. 984--995, 2002.

\bibitem{capon2005high}
J.~Capon, ``High-resolution frequency-wavenumber spectrum analysis,'' \emph{Proceedings of the IEEE}, vol.~57, no.~8, pp. 1408--1418, 2005.

\bibitem{van2002optimum}
H.~L. Van~Trees, \emph{Optimum array processing: Part IV of detection, estimation, and modulation theory}.\hskip 1em plus 0.5em minus 0.4em\relax John Wiley \& Sons, 2002.

\bibitem{johnson2005application}
D.~H. Johnson, ``The application of spectral estimation methods to bearing estimation problems,'' \emph{Proceedings of the IEEE}, vol.~70, no.~9, pp. 1018--1028, 2005.

\bibitem{lacoss1971data}
R.~T. Lacoss, ``Data adaptive spectral analysis methods,'' \emph{Geophysics}, vol.~36, no.~4, pp. 661--675, 1971.

\bibitem{barabell1983improving}
A.~Barabell, ``Improving the resolution performance of eigenstructure-based direction-finding algorithms,'' in \emph{ICASSP'83. IEEE International Conference on Acoustics, Speech, and Signal Processing}, vol.~8.\hskip 1em plus 0.5em minus 0.4em\relax IEEE, 1983, pp. 336--339.

\bibitem{rao2002performance}
B.~D. Rao and K.~S. Hari, ``Performance analysis of root-music,'' \emph{IEEE Transactions on Acoustics, Speech, and Signal Processing}, vol.~37, no.~12, pp. 1939--1949, 2002.

\bibitem{haardt2002unitary}
M.~Haardt and J.~A. Nossek, ``Unitary esprit: How to obtain increased estimation accuracy with a reduced computational burden,'' \emph{IEEE transactions on signal processing}, vol.~43, no.~5, pp. 1232--1242, 2002.

\bibitem{du2002improved}
W.~Du and R.~L. Kirlin, ``Improved spatial smoothing techniques for doa estimation of coherent signals,'' \emph{IEEE Transactions on signal processing}, vol.~39, no.~5, pp. 1208--1210, 2002.

\bibitem{pillai2002forward}
S.~U. Pillai and B.~H. Kwon, ``Forward/backward spatial smoothing techniques for coherent signal identification,'' \emph{IEEE Transactions on acoustics, speech, and signal processing}, vol.~37, no.~1, pp. 8--15, 2002.

\bibitem{stoica2002maximum}
P.~Stoica and K.~C. Sharman, ``Maximum likelihood methods for direction-of-arrival estimation,'' \emph{IEEE Transactions on Acoustics, Speech, and Signal Processing}, vol.~38, no.~7, pp. 1132--1143, 2002.

\bibitem{viberg2002detection}
M.~Viberg, B.~Ottersten, and T.~Kailath, ``Detection and estimation in sensor arrays using weighted subspace fitting,'' \emph{IEEE transactions on Signal Processing}, vol.~39, no.~11, pp. 2436--2449, 2002.

\bibitem{viberg1993exact}
M.~Viberg, P.~Stoica, A.~Nehorai \emph{et~al.}, ``Exact and large sample ml techniques for parameter estimation and detection in array processing,'' in \emph{Radar Array Processing}.\hskip 1em plus 0.5em minus 0.4em\relax Springer-Verlag, 1993.

\bibitem{bohme1985estimation}
J.~B{\"o}hme, ``Estimation of source parameters by approximate maximum likelihood,'' \emph{IFAC Proceedings Volumes}, vol.~18, no.~5, pp. 969--974, 1985.

\bibitem{malioutov2005sparse}
D.~Malioutov, M.~Cetin, and A.~S. Willsky, ``A sparse signal reconstruction perspective for source localization with sensor arrays,'' \emph{IEEE transactions on signal processing}, vol.~53, no.~8, pp. 3010--3022, 2005.

\bibitem{wipf2007empirical}
D.~P. Wipf and B.~D. Rao, ``An empirical bayesian strategy for solving the simultaneous sparse approximation problem,'' \emph{IEEE Transactions on Signal Processing}, vol.~55, no.~7, pp. 3704--3716, 2007.

\bibitem{stoica2010spice}
P.~Stoica, P.~Babu, and J.~Li, ``Spice: A sparse covariance-based estimation method for array processing,'' \emph{IEEE Transactions on Signal Processing}, vol.~59, no.~2, pp. 629--638, 2010.

\bibitem{yang2012off}
Z.~Yang, L.~Xie, and C.~Zhang, ``Off-grid direction of arrival estimation using sparse bayesian inference,'' \emph{IEEE transactions on signal processing}, vol.~61, no.~1, pp. 38--43, 2012.

\bibitem{das2017narrowband}
A.~Das and T.~J. Sejnowski, ``Narrowband and wideband off-grid direction-of-arrival estimation via sparse bayesian learning,'' \emph{IEEE Journal of Oceanic Engineering}, vol.~43, no.~1, pp. 108--118, 2017.

\bibitem{stoica1990performance}
P.~Stoica and A.~Nehorai, ``Performance study of conditional and unconditional direction-of-arrival estimation,'' \emph{IEEE Transactions on Acoustics, Speech, and Signal Processing}, vol.~38, no.~10, pp. 1783--1795, 1990.

\bibitem{ottersten2002performance}
B.~Ottersten, M.~Viberg, and T.~Kailath, ``Performance analysis of the total least squares esprit algorithm,'' \emph{IEEE transactions on signal processing}, vol.~39, no.~5, pp. 1122--1135, 2002.

\bibitem{cramer1999mathematical}
H.~Cram{\'e}r, \emph{Mathematical methods of statistics}.\hskip 1em plus 0.5em minus 0.4em\relax Princeton university press, 1999, vol.~9.

\bibitem{rao1945information}
C.~R. Rao \emph{et~al.}, ``Information and the accuracy attainable in the estimation of statistical parameters,'' \emph{Bull. Calcutta Math. Soc}, vol.~37, no.~3, pp. 81--91, 1945.

\bibitem{wax1992detection}
M.~Wax, ``Detection and localization of multiple sources in noise with unknown covariance,'' \emph{IEEE Transactions on Signal Processing}, vol.~40, no.~1, pp. 245--249, 1992.

\bibitem{valaee2002parametric}
S.~Valaee, B.~Champagne, and P.~Kabal, ``Parametric localization of distributed sources,'' \emph{IEEE Transactions on Signal Processing}, vol.~43, no.~9, pp. 2144--2153, 2002.

\bibitem{hung2002focussing}
H.~Hung and M.~Kaveh, ``Focussing matrices for coherent signal-subspace processing,'' \emph{IEEE Transactions on Acoustics, Speech, and Signal Processing}, vol.~36, no.~8, pp. 1272--1281, 2002.

\bibitem{akaike2003new}
H.~Akaike, ``A new look at the statistical model identification,'' \emph{IEEE transactions on automatic control}, vol.~19, no.~6, pp. 716--723, 2003.

\bibitem{rissanen1978modeling}
J.~Rissanen, ``Modeling by shortest data description,'' \emph{Automatica}, vol.~14, no.~5, pp. 465--471, 1978.

\bibitem{wax2003detection}
M.~Wax and T.~Kailath, ``Detection of signals by information theoretic criteria,'' \emph{IEEE Transactions on acoustics, speech, and signal processing}, vol.~33, no.~2, pp. 387--392, 2003.

\bibitem{vorobyov2003robust}
S.~A. Vorobyov, A.~B. Gershman, and Z.-Q. Luo, ``Robust adaptive beamforming using worst-case performance optimization: A solution to the signal mismatch problem,'' \emph{IEEE transactions on signal processing}, vol.~51, no.~2, pp. 313--324, 2003.

\bibitem{gu2012robust}
Y.~Gu and A.~Leshem, ``Robust adaptive beamforming based on interference covariance matrix reconstruction and steering vector estimation,'' \emph{IEEE Transactions on Signal Processing}, vol.~60, no.~7, pp. 3881--3885, 2012.

\bibitem{paulraj1985estimation}
A.~Paulraj, R.~Roy, and T.~Kailath, ``Estimation of signal parameters via rotational invariance techniques-esprit,'' in \emph{Nineteeth Asilomar Conference on Circuits, Systems and Computers, 1985.}\hskip 1em plus 0.5em minus 0.4em\relax IEEE, 1985, pp. 83--89.

\bibitem{ziskind2002maximum}
I.~Ziskind and M.~Wax, ``Maximum likelihood localization of multiple sources by alternating projection,'' \emph{IEEE Transactions on Acoustics, Speech, and Signal Processing}, vol.~36, no.~10, pp. 1553--1560, 2002.

\bibitem{krim2002two}
H.~Krim and M.~Viberg, ``Two decades of array signal processing research: the parametric approach,'' \emph{IEEE signal processing magazine}, vol.~13, no.~4, pp. 67--94, 2002.

\bibitem{stoica2005spectral}
P.~Stoica, R.~L. Moses \emph{et~al.}, \emph{Spectral analysis of signals}.\hskip 1em plus 0.5em minus 0.4em\relax Pearson Prentice Hall Upper Saddle River, NJ, 2005, vol. 452.

\bibitem{johnson1992array}
D.~H. Johnson and D.~E. Dudgeon, \emph{Array signal processing: concepts and techniques}.\hskip 1em plus 0.5em minus 0.4em\relax Simon \& Schuster, Inc., 1992.

\bibitem{pillai2012array}
S.~U. Pillai, \emph{Array signal processing}.\hskip 1em plus 0.5em minus 0.4em\relax Springer Science \& Business Media, 2012.

\bibitem{ahmed2021deep}
A.~M. Ahmed, U.~S. K.~M. Thanthrige, A.~El~Gamal, and A.~Sezgin, ``Deep learning for doa estimation in mimo radar systems via emulation of large antenna arrays,'' \emph{IEEE Communications Letters}, vol.~25, no.~5, pp. 1559--1563, 2021.

\bibitem{pandey2024grid}
R.~Pandey and S.~Nannuru, ``Grid-free algorithms for direction-of-arrival trajectory localization,'' \emph{The Journal of the Acoustical Society of America}, vol. 155, no.~2, pp. 1379--1390, 2024.

\bibitem{evans1981high}
J.~Evans, ``High resolution angular spectrum estimation technique for terrain scattering analysis and angle of arrival estimation,'' in \emph{1st IEEE ASSP Workshop Spectral Estimat., McMaster Univ., Hamilton, Ont., Canada, 1981}, 1981, pp. 134--139.

\bibitem{friedlander2002direction}
B.~Friedlander and A.~J. Weiss, ``Direction finding using spatial smoothing with interpolated arrays,'' \emph{IEEE Transactions on Aerospace and Electronic Systems}, vol.~28, no.~2, pp. 574--587, 2002.

\bibitem{wang1985coherent}
H.~Wang and M.~Kaveh, ``Coherent signal-subspace processing for the detection and estimation of angles of arrival of multiple wide-band sources,'' \emph{IEEE Transactions on Acoustics, Speech, and Signal Processing}, vol.~33, no.~4, pp. 823--831, 1985.

\bibitem{spielman1986performance}
D.~Spielman, A.~Paulraj, and T.~Kailath, ``Performance analysis of the music algorithm,'' in \emph{ICASSP'86. IEEE International Conference on Acoustics, Speech, and Signal Processing}, vol.~11.\hskip 1em plus 0.5em minus 0.4em\relax IEEE, 1986, pp. 1909--1912.

\bibitem{tran1990cramer}
C.~Tran, ``Cramer-rao bound, music, and maximum likelihood. effects of temporal phase difference,'' Tech. Rep., 1990.

\bibitem{porat1989performance}
B.~Porat and B.~Friedlander, ``Performance analysis of parameter estimation algorithms based on high-order moments,'' \emph{International Journal of adaptive control and signal processing}, vol.~3, no.~3, pp. 191--229, 1989.

\bibitem{liao2016iterative}
B.~Liao, S.-C. Chan, L.~Huang, and C.~Guo, ``Iterative methods for subspace and doa estimation in nonuniform noise,'' \emph{IEEE Transactions on Signal Processing}, vol.~64, no.~12, pp. 3008--3020, 2016.

\bibitem{gupta2003effect}
I.~Gupta and A.~Ksienski, ``Effect of mutual coupling on the performance of adaptive arrays,'' \emph{IEEE Transactions on Antennas and Propagation}, vol.~31, no.~5, pp. 785--791, 2003.

\bibitem{godara2002application}
L.~C. Godara, ``Application of antenna arrays to mobile communications. ii. beam-forming and direction-of-arrival considerations,'' \emph{Proceedings of the IEEE}, vol.~85, no.~8, pp. 1195--1245, 2002.

\bibitem{dibiase2001robust}
J.~H. DiBiase, H.~F. Silverman, and M.~S. Brandstein, ``Robust localization in reverberant rooms,'' in \emph{Microphone arrays: signal processing techniques and applications}.\hskip 1em plus 0.5em minus 0.4em\relax Springer, 2001, pp. 157--180.

\bibitem{benesty2008microphone}
J.~Benesty, J.~Chen, and Y.~Huang, \emph{Microphone array signal processing}.\hskip 1em plus 0.5em minus 0.4em\relax Springer, 2008.

\bibitem{golub2013matrix}
G.~H. Golub and C.~F. Van~Loan, \emph{Matrix computations}.\hskip 1em plus 0.5em minus 0.4em\relax JHU press, 2013.

\bibitem{petersen2008matrix}
K.~B. Petersen, M.~S. Pedersen \emph{et~al.}, ``The matrix cookbook,'' \emph{Technical University of Denmark}, vol.~7, no.~15, p. 510, 2008.

\bibitem{horn2012matrix}
R.~A. Horn and C.~R. Johnson, \emph{Matrix analysis}.\hskip 1em plus 0.5em minus 0.4em\relax Cambridge university press, 2012.

\bibitem{pfister2017discrete}
H.~Pfister, ``Discrete-time signal processing,'' \emph{Lecture Note, pfister. ee. duke. edu/courses/ece485/dtsp. pdf}, 2017.

\bibitem{kay1993fundamentals}
S.~M. Kay, \emph{Fundamentals of statistical signal processing: estimation theory}.\hskip 1em plus 0.5em minus 0.4em\relax Prentice-Hall, Inc., 1993.

\bibitem{poor2013introduction}
H.~V. Poor, \emph{An introduction to signal detection and estimation}.\hskip 1em plus 0.5em minus 0.4em\relax Springer Science \& Business Media, 2013.

\bibitem{boyd2004convex}
S.~P. Boyd and L.~Vandenberghe, \emph{Convex optimization}.\hskip 1em plus 0.5em minus 0.4em\relax Cambridge university press, 2004.

\bibitem{nocedal2006numerical}
J.~Nocedal and S.~J. Wright, \emph{Numerical optimization}.\hskip 1em plus 0.5em minus 0.4em\relax Springer, 2006.

\bibitem{tibshirani1996regression}
R.~Tibshirani, ``Regression shrinkage and selection via the lasso,'' \emph{Journal of the Royal Statistical Society Series B: Statistical Methodology}, vol.~58, no.~1, pp. 267--288, 1996.

\bibitem{harris2020array}
C.~R. Harris, K.~J. Millman, S.~J. Van Der~Walt, R.~Gommers, P.~Virtanen, D.~Cournapeau, E.~Wieser, J.~Taylor, S.~Berg, N.~J. Smith \emph{et~al.}, ``Array programming with numpy,'' \emph{nature}, vol. 585, no. 7825, pp. 357--362, 2020.

\bibitem{virtanen2020scipy}
P.~Virtanen, R.~Gommers, T.~E. Oliphant, M.~Haberland, T.~Reddy, D.~Cournapeau, E.~Burovski, P.~Peterson, W.~Weckesser, J.~Bright \emph{et~al.}, ``Scipy 1.0: fundamental algorithms for scientific computing in python,'' \emph{Nature methods}, vol.~17, no.~3, pp. 261--272, 2020.

\bibitem{hunter2007matplotlib}
J.~D. Hunter, ``Matplotlib: A 2d graphics environment,'' \emph{Computing in science \& engineering}, vol.~9, no.~03, pp. 90--95, 2007.

\bibitem{zheng2024deep}
S.~Zheng, Z.~Yang, W.~Shen, L.~Zhang, J.~Zhu, Z.~Zhao, and X.~Yang, ``Deep learning-based doa estimation,'' \emph{IEEE Transactions on Cognitive Communications and Networking}, vol.~10, no.~3, pp. 819--835, 2024.

\bibitem{salama2025direction}
A.~A. Salama, ``Direction of arrival estimation: A tutorial survey of classical and modern methods,'' \emph{arXiv preprint arXiv:2508.11675}, 2025.

\bibitem{van2002modulation}
H.~L. Van~Trees and E.~Detection, ``Modulation theory, part iv: Optimum array processing,'' \emph{New York: Willey}, 2002.

\bibitem{bartlett1948smoothing}
M.~S. Bartlett, ``Smoothing periodograms from time-series with continuous spectra,'' \emph{Nature}, vol. 161, no. 4096, pp. 686--687, 1948.

\bibitem{pisarenko1973retrieval}
V.~F. Pisarenko, ``The retrieval of harmonics from a covariance function,'' \emph{Geophysical Journal International}, vol.~33, no.~3, pp. 347--366, 1973.

\bibitem{pal2010nested}
P.~Pal and P.~P. Vaidyanathan, ``Nested arrays: A novel approach to array processing with enhanced degrees of freedom,'' \emph{IEEE Transactions on Signal Processing}, vol.~58, no.~8, pp. 4167--4181, 2010.

\bibitem{vaidyanathan2010sparse}
P.~P. Vaidyanathan and P.~Pal, ``Sparse sensing with co-prime samplers and arrays,'' \emph{IEEE Transactions on Signal Processing}, vol.~59, no.~2, pp. 573--586, 2010.

\bibitem{weiss2002effects}
A.~J. Weiss and B.~Friedlander, ``Effects of modeling errors on the resolution threshold of the music algorithm,'' \emph{IEEE Transactions on Signal Processing}, vol.~42, no.~6, pp. 1519--1526, 2002.

\bibitem{van1992azimuth}
A.-J. van~der Veen, P.~B. Ober, and E.~F. Deprettere, ``Azimuth and elevation computation in high resolution doa estimation,'' \emph{IEEE Transactions on Signal Processing}, vol.~40, no.~7, pp. 1828--1832, 1992.

\bibitem{scheibler2018pyroomacoustics}
R.~Scheibler, E.~Bezzam, and I.~Dokmani{\'c}, ``Pyroomacoustics: A python package for audio room simulation and array processing algorithms,'' in \emph{2018 IEEE international conference on acoustics, speech and signal processing (ICASSP)}.\hskip 1em plus 0.5em minus 0.4em\relax IEEE, 2018, pp. 351--355.

\bibitem{salama2016underdetermined}
A.~A. Salama, M.~O. Ahmad, and M.~Swamy, ``Underdetermined doa estimation using mvdr-weighted lasso,'' \emph{Sensors}, vol.~16, no.~9, p. 1549, 2016.

\bibitem{salama2020compressed}
------, ``Compressed sensing doa estimation in the presence of unknown noise,'' \emph{Progress in Electromagnetics Research C}, vol. 102, pp. 47--62, 2020.

\bibitem{salama2017compressive}
------, ``Compressive sensing-based doa estimation using the dantzig selector,'' in \emph{2017 IEEE 60th International Midwest Symposium on Circuits and Systems (MWSCAS)}.\hskip 1em plus 0.5em minus 0.4em\relax IEEE, 2017, pp. 859--862.

\bibitem{salama2021novel}
A.~A. Salama, M.~Morsy, and S.~H. Darwish, ``A novel low complexity cs-based doa estimation technique,'' in \emph{2021 International Telecommunications Conference (ITC-Egypt)}.\hskip 1em plus 0.5em minus 0.4em\relax IEEE, 2021, pp. 1--4.

\bibitem{afsa2017compressive}
A.~A.~S. Afsa, ``Compressive sensing based estimation of direction of arrival in antenna arrays,'' Ph.D. dissertation, Concordia University, 2017.

\bibitem{el2023cramer}
T.~A. El-Rahman, A.~A. Salama, W.~M. Saad, and M.~H. El-Shafey, ``Cramer-rao bound investigation of a novel virtually extended nested arc array for 2d doa estimation,'' \emph{Alexandria Engineering Journal}, vol.~76, pp. 259--274, 2023.

\bibitem{salama2025novel}
A.~A. Salama, A.~K. Awaad, A.~A. Ateya, M.~El~Affendi, S.~Ahmad, A.~Shaalan, and A.~A. Hamdi, ``A novel coherent source doa estimation using adaptive sparse regularization,'' \emph{Engineering, Technology \& Applied Science Research}, vol.~15, no.~4, pp. 25\,660--25\,667, 2025.

\bibitem{el2023crame}
T.~A. El-Rahman, A.~A. Salama, W.~M. Saad, and M.~H. El-Shafey, ``Cramer-rao bound investigation of the double-nested arc array virtual extension and butterfly antenna configuration,'' in \emph{World Conference on Internet of Things: Applications \& Future}.\hskip 1em plus 0.5em minus 0.4em\relax Springer, 2023, pp. 309--331.

\end{thebibliography}
\nocite{*}

\end{document}